%% file: main.tex
\documentclass[dvipsnames]{article}
\usepackage{float}
\usepackage{flafter}
\usepackage{nafuzz_colm2024_conference}

\usepackage[T1]{fontenc}
\usepackage{graphicx}
\usepackage{booktabs}
\usepackage{multirow}
\usepackage{amsmath,amssymb}
\usepackage{enumitem}
\usepackage{listings}

\definecolor{SWElinkblue}{HTML}{2143A5}
\newcommand{\resourcelink}[3]{%
  \noindent\makebox[\linewidth][c]{%
    \makebox[1.45em][c]{\raisebox{-0.18em}{\includegraphics[height=1.15em]{#1}}}%
    \hspace{0.45em}{\bfseries\rmfamily #2:}%
    \hspace{0.45em}\href{#3}{\textcolor{SWElinkblue}{\ttfamily #3}}}%
  \par
}
\newcommand{\vcenterhead}[1]{\raisebox{0.75ex}{\textbf{#1}}}
\newcommand{\preprinttableformat}{%
  \small
  \renewcommand{\arraystretch}{1.08}%
  \setlength{\tabcolsep}{5pt}%
}

\title{SWE-Test: Benchmarking LLM Vulnerability Discovery via Input Prediction}
\input{author_info}
\author{\paperauthors}
\hypersetup{
  pdftitle={SWE-Test: Benchmarking LLM Vulnerability Discovery via Input Prediction},
  pdfauthor={Yuanxiang Shi, Jiayi Lin, Xuanyong Lin, Liangcai Su, Yeheng Duan, Wei Wang, Qi Han, Bing Zhao, Wei Hu, Xander Xu, Chenxiong Qian}
}

\begin{document}

\maketitle

\begingroup
\small
\resourcelink{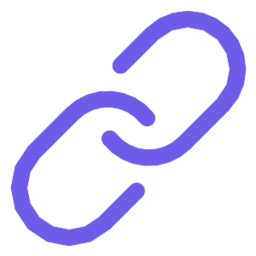}{Website}{https://swe-test-benchmark.github.io/}
\resourcelink{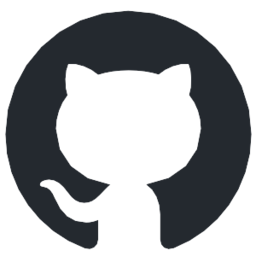}{Dataset}{https://github.com/SWE-Test-Benchmark/SWE-Test-Benchmark}
\endgroup

\begin{abstract}
Vulnerability discovery is becoming an important ability of large language model (LLM) agents: agents that silently miss real defects leave critical software exposed. Rigorously measuring this ability is therefore urgent, but existing benchmarks are gameable through data contamination, score recall against an unknowable vulnerability set, often rely on synthetic bugs, and report a single end-to-end verdict that cannot localize where an agent fails. Vulnerability discovery is a composite ability: an agent must comprehend source code, infer input constraints, construct inputs, execute them, and iteratively correct from feedback. We recast its measurement as an \emph{input-prediction} task with a closed, deterministic ground truth: using coverage-guided fuzzing, we mine deep target branches in real-world C/C++ programs and ask an agent to predict an input that drives execution to a given branch. This decomposes discovery into three task modes over 22 real-world C/C++ programs spanning 15 domains. \emph{Open-loop} and \emph{Feedback-enabled} share 60 fixed-target task instances across 16 of these codebases (13 domains), testing input construction without and with a distance oracle to isolate code comprehension from feedback-driven correction. \emph{Online Arena} instead removes the predefined target and scores path exploration by coverage gain on a separate, partially overlapping pool of 11 programs; agents collectively confirmed 13 distinct bugs across six programs. Evaluating 15 default-effort model-scaffold configurations, the best reaches only 55.0\% pass rate in the Feedback-enabled mode, and the mean across seven paired Claude Code configurations is 36.4\% with feedback versus 19.3\% without. Decomposing failures, we find constraint inference, not navigation, is the dominant bottleneck. We release SWE-Test with a turnkey evaluation environment.

\par\medskip
\noindent\small\textbf{Keywords:} Vulnerability Discovery, LLM Agents,
Fuzzing, Software Testing, \mbox{Benchmark}
\end{abstract}

\section{Introduction}
\label{sec:intro}

Vulnerability discovery has become one of the defining capabilities of frontier coding agents. In early 2026, an AI agent autonomously uncovered a sixteen-year-old vulnerability in FFmpeg, a media library embedded in countless systems, on a line of code that automated fuzzers had executed more than five million times without ever flagging it, and separately chained several Linux kernel flaws into a privilege-escalation exploit~\cite{anthropic2026glasswing}. Anthropic's Frontier Red Team reports that current models can find meaningful zero-day vulnerabilities in well-tested codebases without any specialized harness, adding value on top of the fuzzing infrastructure that security teams have refined for years~\cite{anthropic2026zerodayrisk}. As agents move from writing code to auditing it, their capacity to surface security-relevant defects bears directly on the security of the software supply chain, which makes measuring that capacity, rigorously and at scale, an urgent problem.

Unfortunately, existing benchmarks used to measure it rest on assumptions that do not hold for real-world vulnerability discovery. First, security benchmarks built from public records can be gamed: their tasks are absorbed into model training and solved partly from memory. A recent analysis finds that on InterCode-CTF a frontier model has a solution-leakage rate of about 14\%, attributed to the benchmark's age and likely presence in the training corpus~\cite{peng2026hackershallucinatorscomprehensiveanalysis}; newer security benchmarks try to dodge this by using only post-cutoff challenges (CyBench, 2022 to 2024~\cite{zhang2025cybenchframeworkevaluatingcybersecurity}; CVE-Bench, May to June 2024~\cite{zhu2025cvebenchbenchmarkaiagents}), but that only delays the problem as cutoffs advance. Second, the set of vulnerabilities in a codebase is unknown and cannot be enumerated, so recall against a fixed list of known vulnerabilities is a biased metric. CyberGym, for example, asks the agent to reproduce one specific disclosed vulnerability and credits only that target~\cite{wang2026cybergymevaluatingaiagents}, so an agent that finds a different genuine defect scores nothing and the reported recall is not a meaningful fraction of an unknowable whole. Third, some benchmarks sidestep these difficulties with synthetic bugs, which are easy to grade but rarely match the constraint structure and cross-procedure depth of real defects.

We observe that vulnerability discovery is a composite ability: an agent must comprehend source code, infer input constraints, construct inputs, execute them, and iteratively correct its approach from feedback. Building on this, we recast the measurement of an agent's vulnerability-discovery ability as an
\emph{input-prediction} task, decomposing it into three underlying abilities: deep
code comprehension, feedback-driven correction, and path exploration. To probe the first two, we use fuzzing to mine, from real programs, deep target branches together with the execution paths that reach them, and we ask the agent to generate an input that drives execution to a given branch. Reaching such a branch requires the agent to understand deep code, since the path from the program entry to the target typically spans multiple files and functions, and to reverse-reason from that target back to a satisfying input. Beyond the source, we then add an environment feedback signal that reports the agent's approximate distance to the target, so the agent can continuously adjust its candidate and converge on an input that reaches the branch. Finally, to probe the third ability, we place the agent in a fuzzing process that never stops on its own and turn the single target branch into \emph{all} branches: the agent now aims to maximize code coverage, with coverage itself serving as the feedback signal, and repeatedly generates new inputs to reach new branches by reasoning over the current coverage and the source. This formulation is hard to game, since the agent must synthesize an input that satisfies a branch identified by line numbers in a specific build, for which there is no public answer to memorize, and it needs no predefined vulnerability list: posed over real programs, it can trigger real bugs.

Building on this design, we introduce \textbf{SWE-Test}, a benchmark with three modes. The \emph{Open-loop} mode exposes deep code comprehension alone, providing only source and a target branch. The \emph{Feedback-enabled} mode adds the distance oracle, exercising feedback-driven correction on the same targets. The \emph{Online Arena} mode removes the target entirely and scores coverage gain under the never-stopping loop above. The Open-loop and Feedback-enabled modes span 13 domains and 16 real-world programs, including language runtimes, parsers, networking, font and text shaping, XML, cryptography, and compilers. In Online Arena, we run agents continuously for 12 hours.

Across 15 default-effort model-scaffold configurations, the best feedback-enabled pass rate is 55.0\%. Among the seven Claude Code configurations with 60 valid trials in both modes, feedback raises mean pass rate from 19.3\% to 36.4\%, target-function reach from 62.6\% to 75.2\%, and near-to-pass conversion from 30.8\% to 48.4\%. In Online Arena, the five evaluated configurations produce 18 model-bug detections, or 13 confirmed bugs across six programs after deduplication by upstream issue or fix.

To make these measurements reproducible and easy to adopt, we release SWE-Test with a standardized, Harbor-based turnkey deployment and evaluation environment that provisions the containerized tasks, runs any supported model-scaffold configuration, and reports per-task and aggregate scores.

\section{Related Work}
\label{sec:related}

Existing vulnerability-discovery benchmarks span several increasingly realistic settings, from reproducing disclosed vulnerabilities to open-ended discovery in real software. Table~\ref{tab:related-benchmarks} compares CyberGym~\cite{wang2026cybergymevaluatingaiagents}, CVE-Bench~\cite{zhu2025cvebenchbenchmarkaiagents}, CyBench~\cite{zhang2025cybenchframeworkevaluatingcybersecurity}, BountyBench~\cite{zhang2025bountybench}, AgentCyberRange~\cite{liu2026agentcyberrange}, and SWE-Test. A dash indicates that the feature is not the benchmark's primary objective or is not reported in the available description.

\begin{table}[H]
\centering
\caption{Comparison with selected vulnerability-discovery benchmarks.}
\label{tab:related-benchmarks}
\preprinttableformat
\setlength{\tabcolsep}{3pt}
\begin{tabular}{@{}p{0.23\columnwidth}ccccccc@{}}
\toprule
\vcenterhead{Benchmark} & \shortstack{Real\\software} & \shortstack{Predefined\\target} & \shortstack{Open\\discovery} & \shortstack{Input\\grading} & \shortstack{Controlled\\feedback} & \shortstack{Anti-\\contam.} & \shortstack{Long\\horizon} \\
\midrule
CyberGym & \checkmark & \checkmark & -- & -- & \checkmark & -- & -- \\
CVE-Bench & \checkmark & \checkmark & -- & -- & \checkmark & -- & -- \\
CyBench & -- & \checkmark & -- & -- & \checkmark & -- & -- \\
BountyBench & \checkmark & -- & \checkmark & -- & \checkmark & -- & -- \\
AgentCyberRange & \checkmark & -- & \checkmark & -- & \checkmark & -- & -- \\
SWE-Test & \checkmark & \checkmark & \checkmark & \checkmark & \checkmark & \checkmark & \checkmark \\
\bottomrule
\end{tabular}
\end{table}

The comparison highlights three recurring limitations. First, many CTF and issue-driven benchmarks rely on public, fixed tasks, making their results vulnerable to contamination and memorization. Live variants delay this risk rather than eliminate it. Transformed or synthetic tasks improve scalability and reduce direct memorization, but may not preserve the control-flow depth or input constraints of real multi-file software. Second, CVE- and PoC-oriented benchmarks evaluate the reproduction or exploitation of a predefined vulnerability. Although this design provides a clear execution target, it does not credit different defects discovered by an agent. Recall over known vulnerabilities cannot represent coverage of an unknowable vulnerability set. Third, end-to-end penetration-testing benchmarks combine discovery, exploitation, and post-exploitation into one outcome. This aggregation obscures whether failure arises from code comprehension, input construction, feedback use, or target selection.

SWE-Test addresses these limitations by reframing vulnerability discovery as an \textbf{input-prediction} task. The agent receives the source of a real C/C++ program and must construct an input that exercises a specified decision outcome in the code. This formulation decomposes vulnerability-discovery ability into \textbf{code comprehension}, \textbf{feedback-driven correction}, and \textbf{path exploration}. Each target originates from an execution observed during prior automated testing, with a hidden input establishing both reachability and deterministic ground truth. This closed objective avoids evaluating recall against an inherently incomplete set of known vulnerabilities. Tying targets to specific builds reduces contamination, while withholding satisfying inputs limits answer memorization. The graded reward distinguishes failure to reach the target function from failure to satisfy the target branch, enabling precise diagnosis on real software with realistic input constraints.

\section{SWE-Test Benchmark}
\label{sec:design}

This section presents SWE-Test as a concrete realization of the input-prediction formulation introduced in Section~\ref{sec:intro}. Section~\ref{sec:bench:overview} summarizes the dataset and its application domains. Section~\ref{sec:bench:modes} defines the three target abilities, their corresponding task modes, and the associated sandbox and reward design. Section~\ref{sec:bench:construction} describes how reachable targets are mined from real fuzzing executions and converted into contamination-resistant benchmark tasks.

\subsection{Overview}
\label{sec:bench:overview}

\begin{figure}[t]
\centering
\includegraphics[width=0.85\linewidth]{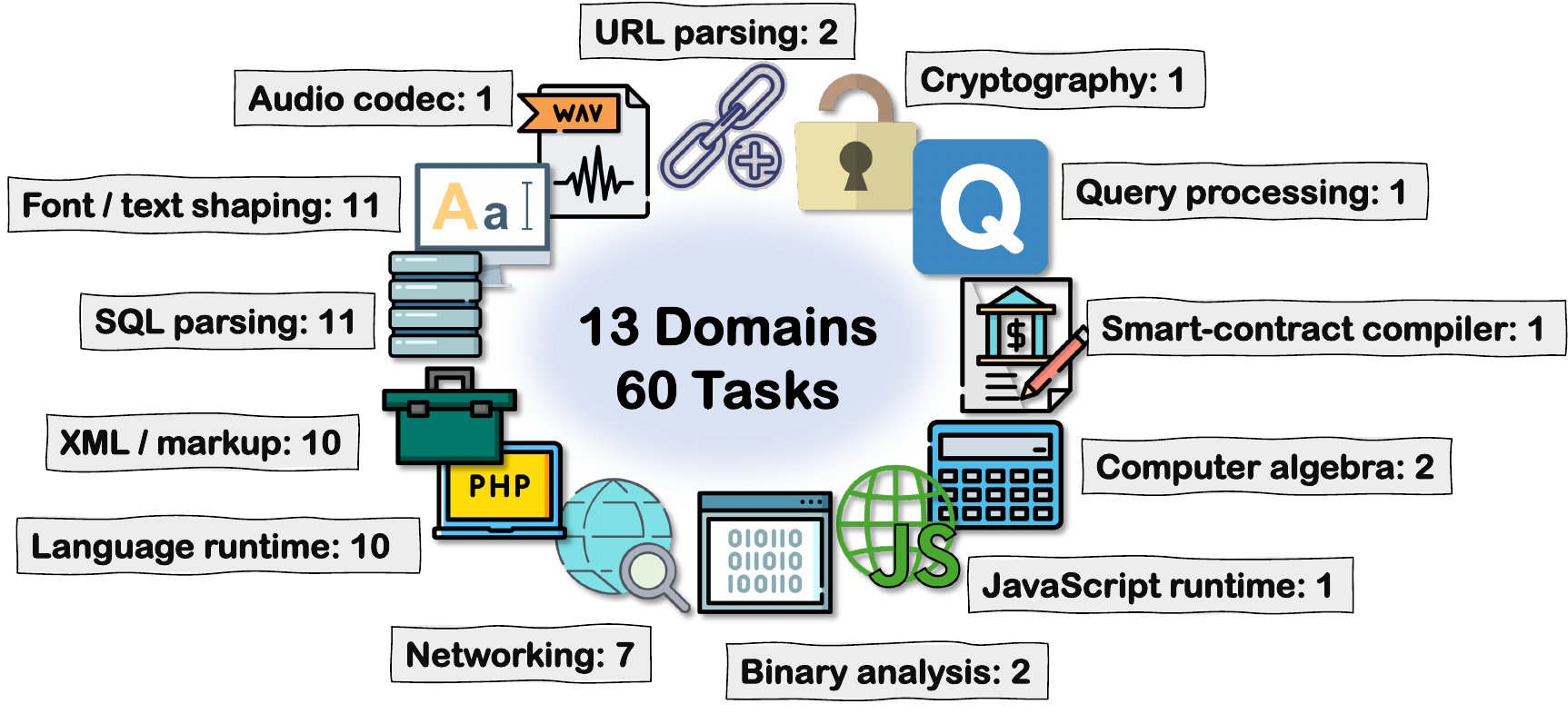}
\caption{Composition of SWE-Test's 60 fixed-target input-prediction tasks (Open-loop and Feedback-enabled modes), covering 16 real-world C/C++ programs across 13 application domains.}
\label{fig:dataset}
\end{figure}

\paragraph{Dataset.}
The fixed-target pool used by Open-loop and Feedback-enabled comprises 60 input-prediction tasks from 16 programs across 13 domains; Online Arena uses a separate, partially overlapping pool, detailed below. Figure~\ref{fig:dataset} summarizes the fixed-target distribution, spanning parsers, network protocols, text and binary formats, language runtimes, cryptography, and compilers.

SWE-Test evaluates input prediction through three complementary modes. \emph{Open-loop} requires the agent to construct an input for a fixed target branch using source code without runtime feedback, emphasizing code comprehension. \emph{Feedback-enabled} retains the same target but provides a distance oracle, allowing the agent to refine candidate inputs through feedback. \emph{Online Arena} removes the predefined target and requires the agent to select unexplored code paths using coverage information, thereby evaluating self-directed path exploration.

Open-loop and Feedback-enabled draw their 60 tasks from the same fixed-target pool of 16 programs across 13 domains. Online Arena runs on a separate pool of 11 programs, since it needs a fuzzer-saturated corpus rather than a single mined branch (Section~\ref{sec:bench:construction}); five of these (HarfBuzz, libxml2, OpenSSL, PHP, QuickJS) overlap with the fixed-target pool, and six (CPython, libjpeg, libpcap, libpng, Lua, SQLite) are exclusive to Online Arena, adding image-codec (libjpeg, libpng) and database-engine (SQLite) domains not present in the fixed-target pool. In total, the benchmark spans 22 distinct C/C++ codebases across 15 domains.

\subsection{Task Modes}
\label{sec:bench:modes}
\label{sec:bench:task}

\begin{figure}[t]
\centering
\includegraphics[width=\columnwidth]{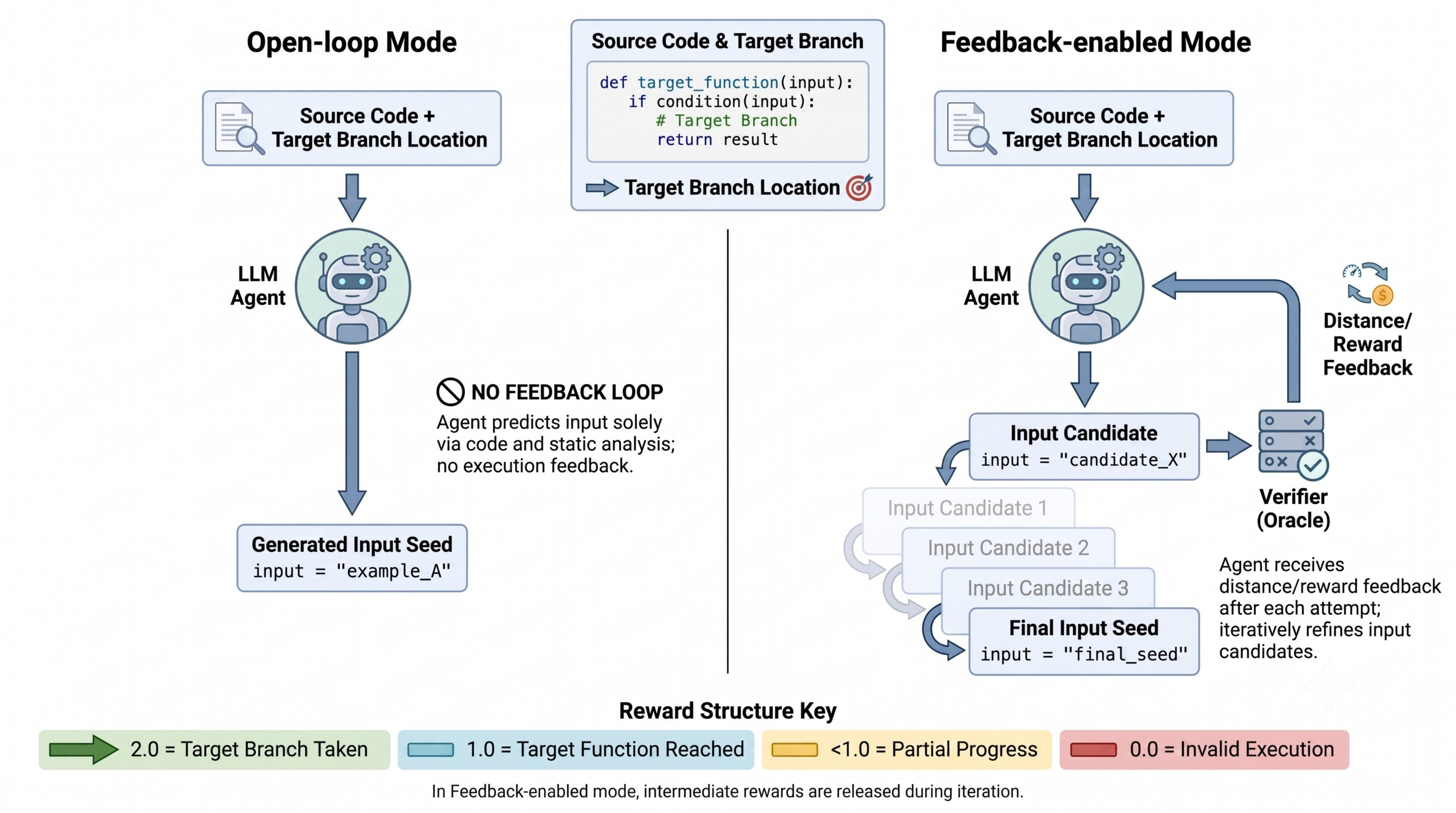}
\caption{Task example contrasting Open-loop and Feedback-enabled modes on a shared target branch.}
\label{fig:task-example-default-feedback}
\end{figure}

\paragraph{Open-loop mode.}
The agent is given the target program's source and a single branch condition that is currently \emph{not} taken, and must produce an input that takes it. Figure~\ref{fig:task-example-default-feedback} illustrates this task interface alongside its feedback-enabled counterpart. No execution oracle is provided: the agent cannot observe whether any candidate input moves it closer to the target. Success therefore requires the agent to trace data flow and construct a satisfying input by reasoning alone, a capability adjacent to static constraint inference and symbolic execution, performed mentally over the source.

The reward is $2.0$ when execution takes the target branch and $1.0$ when it reaches the containing function without satisfying the branch condition. If execution does not reach the target function, it receives a trace-progress reward between $0.0$ and $1.0$, based on its overlap with the ground-truth execution path. A reward of $0.0$ indicates no path overlap.

\paragraph{Feedback-enabled mode.}
On the \emph{same} targets, the agent may additionally run a verification oracle that reports how close the current input is to triggering the branch. Figure~\ref{fig:task-example-default-feedback} shows the shared target and the additional verifier interface. This admits the construct-execute-observe-revise loop. Because the only difference from the Open-loop mode is the presence of feedback, the gap between the two modes isolates feedback-driven correction while holding comprehension fixed.

The feedback-enabled mode uses the same reward levels and exposes the intermediate reward after each verifier query.

\begin{figure}[t]
\centering
\includegraphics[width=\columnwidth]{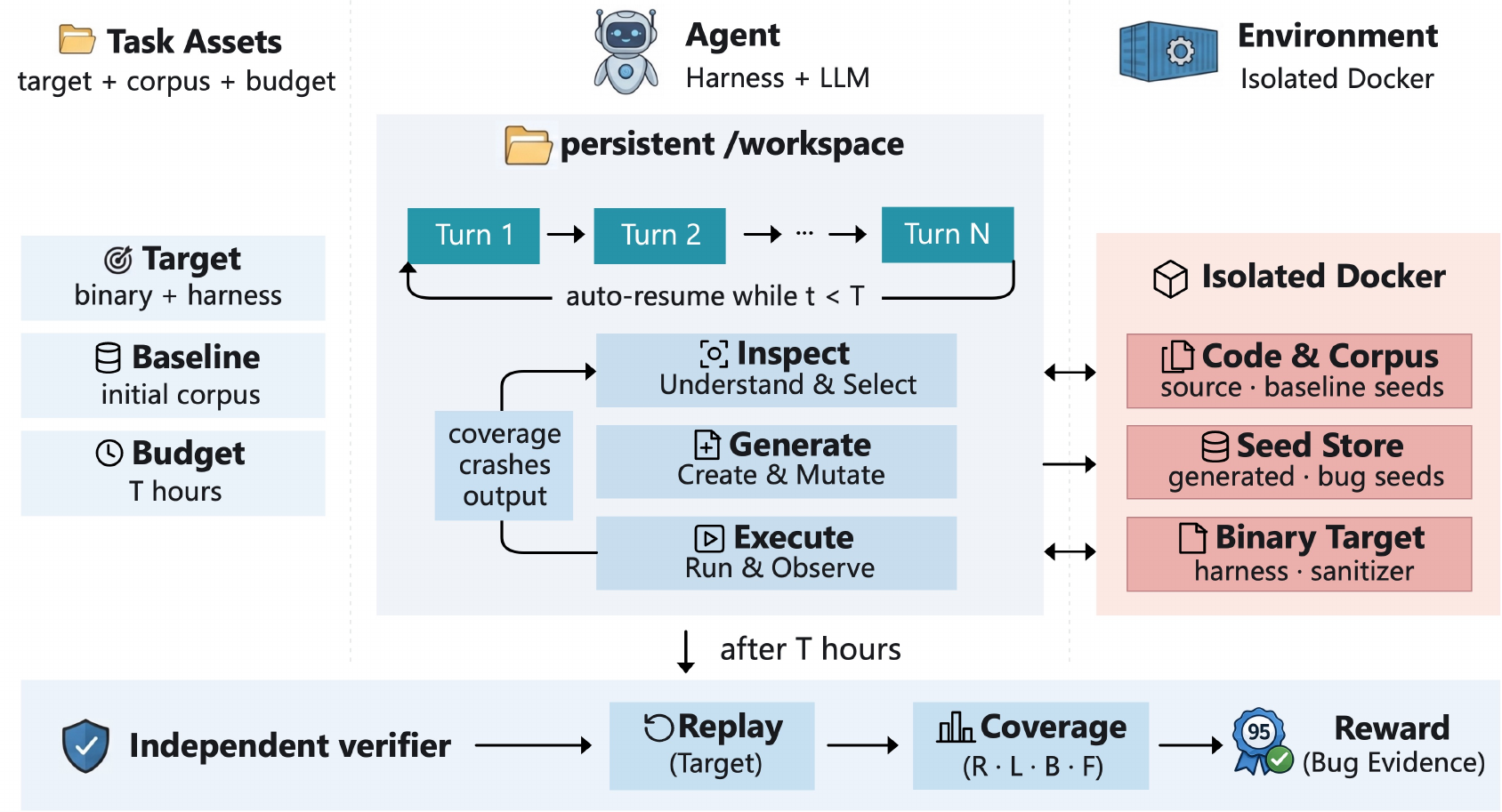}
\caption{Online Arena execution and verification pipeline.}
\label{fig:online-running-pipeline}
\end{figure}

\paragraph{Online Arena.}
No target branch is specified at all, so the distance oracle of the feedback-enabled mode no longer applies. Instead, the agent is given a coverage tool that reports the current coverage and the list of still-uncovered branches, and runs in a long-horizon loop: it queries coverage, reads per-line annotated source to find uncovered branches worth attacking, generates new inputs, and re-measures. Figure~\ref{fig:online-running-pipeline} summarizes this multi-turn loop, its isolated execution environment, and the independent verification stage. This is still feedback, but of a different kind, namely undirected coverage feedback rather than distance to a fixed target. The open-ended objective is broader than coverage: while reading uncovered code, the agent also hunts for defects, including crashes, hangs, state corruption, leaks, and functional or semantic misbehavior, and routes seeds associated with confirmed findings to a separate corpus. Coverage gain rather than a single branch flip remains the quantitative score, since the set of real vulnerabilities in a codebase is unknowable and recall against a fixed bug list would reintroduce the bias this benchmark avoids; discovered bugs are logged as artifacts. The step up from the feedback-enabled mode isolates the added value of self-directed exploration: the agent must now decide \emph{what} to pursue, not merely \emph{how} to reach a given target.

The Online Arena reward is the mean normalized gain in regions, lines, branches, and functions relative to the baseline corpus; confirmed bugs are reported separately. For each coverage dimension $d$, we compute
\[
  \mathit{reward}_d = \mathrm{clamp}\!\left(\frac{\mathit{final}_d - \mathit{baseline}_d}{100 - \mathit{baseline}_d},\, 0,\, 1\right),
\]
and report the mean of the four $\mathit{reward}_d$ values.

\subsection{Benchmark Construction}
\label{sec:bench:construction}

\begin{figure}[t]
\centering
\includegraphics[width=\textwidth]{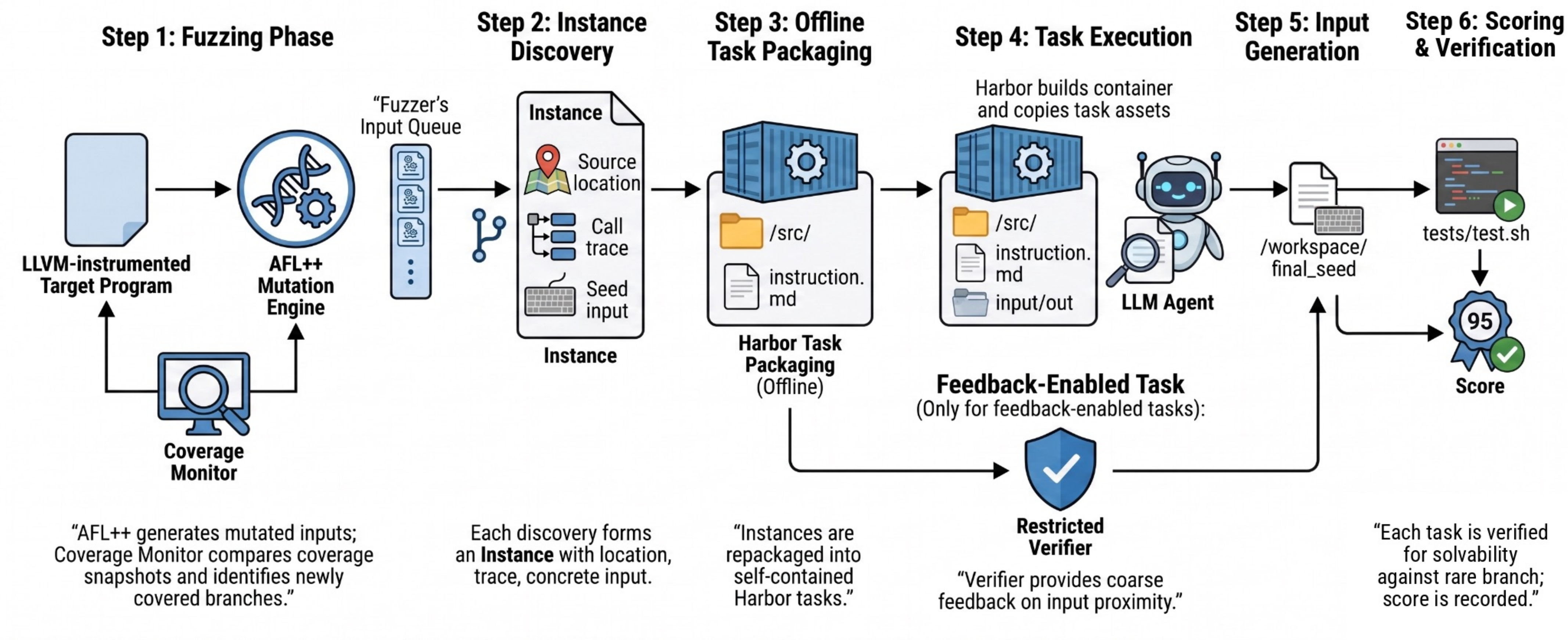}
\caption{Construction pipeline for Open-loop and Feedback-enabled tasks.}
\label{fig:benchmark-construction-pipeline}
\end{figure}

SWE-Test tasks are not hand-authored. All three modes are built from the same coverage-guided fuzzing runs, which guarantees that every target is real and reachable, but the two kinds of task consume those runs differently.

\paragraph{Open-loop and Feedback-enabled tasks.}
Figure~\ref{fig:benchmark-construction-pipeline} summarizes the six-stage construction and evaluation pipeline. We instrument each target for LLVM source-based coverage, use AFL++~\cite{fioraldi2020aflplusplus} to generate candidate inputs, and monitor successive coverage snapshots. When an input reaches a previously uncovered branch, we record the branch location, call trace, and concrete input as a task instance. An offline stage packages each instance with the program source and instructions in a self-contained Harbor task. During evaluation, Harbor builds an isolated container in which the agent must write its candidate to \texttt{/workspace/final\_seed}. Feedback-enabled tasks expose a restricted verifier that returns only coarse progress, whereas Open-loop tasks provide no runtime feedback. After the agent finishes, the root-owned \texttt{tests/test.sh} evaluates the final input and records its score. The hidden input discovered during construction establishes that every target branch is reachable, while the newly covered status selects branches that were absent from the preceding corpus.

\paragraph{Online Arena tasks.}
This mode needs no single branch, but it does need a starting point that is hard. We take the corpus a fuzzer has accumulated after 72 hours on a target and hand it to the agent as its starting seeds. The reason is how coverage grows: it rises steeply and then flattens, so the easily reachable branches are, empirically, already covered well within 72 hours. Starting from such a saturated corpus puts every agent at the same hard frontier, where the branches that remain are precisely those blind mutation could not reach. This reduces the agent budget spent rediscovering easy coverage and focuses the comparison on the harder frontier, because branches readily reached by fuzzing provide limited evidence for distinguishing agents' path-exploration ability.

\section{Experimental}
\label{sec:setup}

\subsection{Experimental Setup and Research Questions}
\label{sec:setup:rqs}
\label{sec:setup:models}
\label{sec:setup:protocol}

\begin{itemize}[nosep]
    \item \textbf{RQ1 (Leaderboard).} Which model/scaffold configuration
exhibits the strongest vulnerability-discovery ability, and how large is the
performance gap across configurations? (Section~\ref{sec:results:rq1})
    \item \textbf{RQ2 (Bottleneck diagnosis).} Why do agents fail to translate
their code-comprehension ability into successful task completion, and how do
these failure patterns vary across modes and domains?
(Section~\ref{sec:results:rq2})
    \item \textbf{RQ3 (Evaluation sensitivity).} How sensitive are measured
vulnerability-discovery results to reasoning effort and scaffold choice?
(Section~\ref{sec:ablation})
\end{itemize}

\paragraph{Models, scaffolds, and protocol.}
A \emph{configuration} pairs a language model with an agent \emph{scaffold} managing tool use, context, and permissions. We evaluate vendor-specific scaffolds, namely Codex, Kimi Code, and Qwen Coder, and generic scaffolds, namely Claude Code and Terminus~2; unless otherwise noted, controlled comparisons use a fixed scaffold, with Terminus~2 treated as an ablation (Section~\ref{sec:ablation}). We evaluate seven models from four families: DeepSeek-V4-Pro from DeepSeek~\cite{deepseek2026v4pro}; GLM-5.1 and GLM-5.2 from GLM~\cite{zai2026glm51,zai2026glm52}; Kimi-K2.6 and Kimi-K3 from Kimi~\cite{kimiwebsite}; and Qwen3.7-Max and Qwen3.8-Max from Qwen~\cite{qwen2026textgenerationmodels}. All tasks run in isolated Harbor containers; Open-loop and Feedback-enabled tasks receive a 24-hour budget, one CPU, 8\,GB memory, and network access, scoring runs separately, exposing only the numeric reward. Models use default reasoning effort unless otherwise noted (Section~\ref{sec:ablation:effort}), and each configuration receives one trial per task. Pass rates use only valid trials, excluding empty scores. Table~\ref{tab:leaderboard} summarizes the evaluated configurations and pass rates in both fixed-target modes.

\begin{table}[ht]
\centering
\caption{Feedback-enabled/Open-loop pass rates by model and scaffold.}
\label{tab:leaderboard}
\preprinttableformat
\setlength{\tabcolsep}{3pt}
\begin{tabular}{@{}lrrrrr@{}}
\toprule
& \textbf{Claude Code} & \textbf{Terminus 2} & \textbf{Codex} & \textbf{Kimi Code} & \textbf{Qwen Coder} \\
\midrule
Qwen3.8-Max      & \textbf{55.0}/\textbf{50.0}\% & --        & --          & --          & -- \\
DeepSeek-V4-Pro  & 35.0/8.3\%                    & 20.7/2.0\% & --        & --          & -- \\
GLM-5.1          & 23.3/1.7\%                    & 1.7/1.7\%  & --        & --          & -- \\
GLM-5.2          & \underline{41.7}/\underline{26.7}\% & 10.0/6.9\% & -- & --          & -- \\
Kimi-K2.6        & 35.0/8.3\%                    & 3.7/3.6\%  & --        & 16.7/3.8\%  & -- \\
Kimi-K3          & 35.0/25.0\%                   & --        & --        & --          & -- \\
Qwen3.7-Max      & 30.0/15.0\%                   & 0.0/1.7\%  & 1.7/3.3\% & --          & 13.3/6.7\% \\
\midrule
\textbf{Mean}    & \textbf{36.4}/\textbf{19.3}\% & 7.2/3.2\%  & 1.7/3.3\% & 16.7/3.8\%  & 13.3/6.7\% \\
\bottomrule
\end{tabular}
\end{table}

\subsection{Which Configuration Performs Best?}
\label{sec:results:rq1}

The clearest sign of vulnerability-discovery capability is whether an agent finds real, confirmed bugs given time to explore, not whether it flips a single pre-selected branch under a fixed evaluation budget. We evaluate using three complementary signals: bug count and normalized coverage reward under an extended Online Arena budget, and pass rate under the standard feedback-enabled mode. Tables~\ref{tab:rq1-online} and~\ref{tab:rq1-bugs} report the two Online Arena signals before the controlled fixed-target comparison.

\begin{table}[H]
\centering
\caption{Normalized coverage reward in 12-hour Online mode.}
\label{tab:rq1-online}
\preprinttableformat
\begin{tabular}{@{}lrrrrr@{}}
\toprule
\vcenterhead{Program} & \shortstack{\textbf{Qwen3.8-}\\\textbf{Max-Preview}} & \shortstack{\textbf{Kimi-}\\\textbf{K2.6}} & \shortstack{\textbf{GLM-}\\\textbf{5.2}} & \shortstack{\textbf{DeepSeek-}\\\textbf{V4-Pro}} & \shortstack{\textbf{Qwen3.7-}\\\textbf{Max}} \\
\midrule
CPython  & 0.22 & 0.21 & 0.23 & 0.33 & 0.19 \\
HarfBuzz & 0.22 & 0.11 & 0.19 & 0.20 & 0.18 \\
libjpeg  & 0.14 & 0.14 & 0.14 & 0.22 & 0.14 \\
libpcap  & 0.60 & 0.60 & 0.59 & 0.58 & 0.56 \\
libpng   & 0.27 & 0.27 & 0.27 & 0.26 & 0.27 \\
libxml2  & 0.19 & 0.19 & 0.19 & 0.16 & 0.18 \\
Lua      & 0.30 & 0.30 & 0.23 & 0.20 & 0.23 \\
OpenSSL  & 0.15 & 0.17 & 0.16 & 0.14 & 0.14 \\
PHP      & 0.16 & 0.17 & 0.13 & 0.08 & 0.07 \\
QuickJS  & 0.49 & 0.42 & 0.36 & 0.35 & 0.22 \\
SQLite   & 0.71 & 0.69 & 0.70 & 0.58 & 0.57 \\
\midrule
\textbf{Mean} & \textbf{0.3136} & 0.2977 & 0.2900 & 0.2818 & 0.2500 \\
\bottomrule
\end{tabular}
\end{table}

\begin{table}[H]
\centering
\caption{Per-model counts of confirmed bugs from 12-hour Online runs.}
\label{tab:rq1-bugs}
\preprinttableformat
\begin{tabular}{@{}lrrrrr@{}}
\toprule
\vcenterhead{Program} & \shortstack{\textbf{Qwen3.8-}\\\textbf{Max-Preview}} & \shortstack{\textbf{Kimi-}\\\textbf{K2.6}} & \shortstack{\textbf{GLM-}\\\textbf{5.2}} & \shortstack{\textbf{DeepSeek-}\\\textbf{V4-Pro}} & \shortstack{\textbf{Qwen3.7-}\\\textbf{Max}} \\
\midrule
CPython  & 2 & 0 & 0 & 0 & 0 \\
HarfBuzz & 1 & 1 & 0 & 1 & 0 \\
libjpeg  & 0 & 0 & 0 & 0 & 0 \\
libpcap  & 1 & 0 & 0 & 0 & 0 \\
libpng   & 0 & 0 & 1 & 0 & 0 \\
libxml2  & 0 & 0 & 0 & 0 & 0 \\
Lua      & 0 & 0 & 0 & 0 & 0 \\
OpenSSL  & 0 & 0 & 0 & 0 & 0 \\
PHP      & 0 & 0 & 0 & 0 & 0 \\
QuickJS  & 3 & 3 & 2 & 1 & 0 \\
SQLite   & 0 & 0 & 2 & 0 & 0 \\
\midrule
\textbf{Total} & \textbf{7} & \textbf{4} & \textbf{5} & \textbf{2} & \textbf{0} \\
\bottomrule
\end{tabular}
\end{table}

\paragraph{Bug counts under the Online Arena.}
For the five configurations evaluated in Online Arena, we extend the per-program budget to 12 hours of wall-clock time and count the manually confirmed bug identities discovered by each model. Table~\ref{tab:rq1-bugs} reports the result.
We evaluate five models in 12-hour Online runs: Qwen3.8-Max-Preview, Kimi-K2.6,
GLM-5.2, DeepSeek-V4-Pro, and Qwen3.7-Max. We manually validate the reported
findings.
Table~\ref{tab:rq1-bugs} reports per-model counts: a bug independently found by
multiple models contributes once to each corresponding model column. Across all
five models, the 18 model-bug detections reduce to 13 distinct bugs in six
programs after matching duplicate findings by their upstream issue or fix.

Running every configuration for 12 hours per program is expensive, so as a
complementary signal we also report normalized composite coverage reward from the
same 12-hour Online runs in Table~\ref{tab:rq1-online}. Qwen3.8-Max-Preview leads
both summaries, with seven confirmed bugs and a mean coverage reward of 0.3136.
Overall, the bug counts broadly follow the coverage-reward trend; Kimi-K2.6 and
GLM-5.2 exchange positions with only a small difference in mean coverage reward
(0.2977 vs.\ 0.2900).
\paragraph{Leaderboard under the feedback-enabled mode.}
Online Arena is expensive to run at scale, so we also compare the seven Claude Code configurations with 60 valid trials in both modes. Table~\ref{tab:leaderboard} summarizes the evaluated configurations and pass rates in both fixed-target modes. Qwen3.8-Max leads the Feedback-enabled mode at 55.0\%, followed by GLM-5.2 at 41.7\%, and DeepSeek-V4-Pro, Kimi-K2.6, and Kimi-K3 at 35.0\%.

\paragraph{Why feedback matters.}
Across the seven paired Claude Code configurations, feedback raises mean pass rate from 19.3\% to 36.4\% and improves every configuration. The corresponding gains in target-function reach and near-to-pass conversion indicate that feedback supports both code-path navigation and branch-constraint refinement.

\paragraph{Answer to RQ1.}
Under a 12-hour Online Arena budget, Qwen3.8-Max-Preview records the most
confirmed bugs among the five evaluated models (seven) and the highest mean
coverage reward (0.3136). Across models, these runs yield 13 distinct confirmed
bugs in six programs after deduplication. In the separate feedback-enabled
evaluation, Qwen3.8-Max\,+\,Claude Code leads at 55.0\%. RQ2 next asks what
separates the leaders from the rest.

\subsection{Why Do Agents Fail?}
\label{sec:results:rq2}

We bucket every trial into a failure stratum using the oracle's reward: \textbf{navigation failure} ($r{<}1.0$, never reached the target function; this includes hard failures, no-op resubmissions of the seed, and partial credit for how far the call trace progressed toward the function), \textbf{constraint-satisfaction failure} ($r{=}1.0$, reached the function but did not satisfy the target branch constraint), and \textbf{pass} ($r{=}2.0$).

\paragraph{Failure is dominated by constraint satisfaction, not navigation.}
Table~\ref{tab:rq2-strata} reports three outcomes for the seven Claude Code configurations with 60 valid trials in both modes. Under feedback, constraint-satisfaction failures generally outnumber navigation failures: agents reach the right function more often than they satisfy its target branch constraint. The bottleneck is therefore not only locating the relevant code but satisfying the constraint that guards the target branch once found.

\begin{table}[t]
\centering
\caption{Outcome counts under Claude Code on identical targets ($n=60$).}
\label{tab:rq2-strata}
\preprinttableformat
\begin{tabular}{@{}lrrr@{}}
\toprule
\textbf{Model} & \textbf{Nav.} & \textbf{Constr.} & \textbf{Pass} \\
\midrule
\multicolumn{4}{@{}l}{\emph{Feedback-enabled}}\\
Qwen3.8-Max      & 10 & 17 & 33 \\
DeepSeek-V4-Pro  & 12 & 27 & 21 \\
GLM-5.1          & 11 & 35 & 14 \\
GLM-5.2          & 10 & 25 & 25 \\
Kimi-K2.6        & 19 & 20 & 21 \\
Kimi-K3          & 29 & 10 & 21 \\
Qwen3.7-Max      & 13 & 29 & 18 \\
\midrule
\multicolumn{4}{@{}l}{\emph{Open-loop}}\\
Qwen3.8-Max      & 13 & 17 & 30 \\
DeepSeek-V4-Pro  & 29 & 26 & 5 \\
GLM-5.1          & 30 & 29 & 1 \\
GLM-5.2          & 23 & 21 & 16 \\
Kimi-K2.6        & 26 & 29 & 5 \\
Kimi-K3          & 18 & 27 & 15 \\
Qwen3.7-Max      & 18 & 33 & 9 \\
\bottomrule
\end{tabular}
\end{table}

\begin{figure}[H]
\centering
\includegraphics[width=0.70\linewidth]{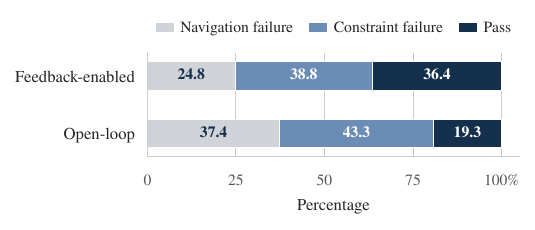}
\caption{Navigation and target-constraint satisfaction by mode.}
\label{fig:rq2-conversion}
\end{figure}

\paragraph{Feedback improves both stages.}
Among trials that reached the target function, the fraction that converted to a pass (near-to-pass conversion) is 48.4\% with the verifier and 30.8\% without it. Feedback also increases the fraction of trials that reach the target function, from 62.6\% to 75.2\%. It therefore improves both navigation and target-constraint satisfaction rather than acting on only one stage. Figure~\ref{fig:rq2-conversion} visualizes the outcome distributions.

\paragraph{Domain modulates the bottleneck.}
The dominant failure mode also varies by task family (Table~\ref{tab:rq2-domain}, which pools the Feedback-enabled trials of the seven paired Claude Code configurations). SQL-parser tasks show 82\% navigation failure, whereas PHP tasks show 17\%. These differences suggest that input structure and source-to-harness mapping influence whether agents reach the target before solving its branch constraint.

\begin{table}[H]
\centering
\caption{Failure outcomes by task family (paired Claude Code configurations).}
\label{tab:rq2-domain}
\preprinttableformat
\renewcommand{\arraystretch}{0.96}
\setlength{\tabcolsep}{9pt}
\begin{tabular}{@{}lrrr@{}}
\toprule
\vcenterhead{Task family} & \shortstack{\textbf{Nav.}\\\textbf{fail (\%)}} & \shortstack{\textbf{Constr.}\\\textbf{fail (\%)}} & \shortstack{\textbf{Pass}\\\textbf{(\%)}} \\
\midrule
Ada URL parsing                & 0  & 100 & 0 \\
Bloaty                         & 14 & 86  & 0 \\
cURL                           & 14 & 14  & 72 \\
FreeType                       & 14 & 29  & 57 \\
HarfBuzz                       & 16 & 77  & 7 \\
JQ                             & 0  & 100 & 0 \\
libxml2                        & 6  & 33  & 61 \\
mBedTLS                        & 0  & 0   & 100 \\
OpenSSL                        & 14 & 14  & 72 \\
OpenThread                     & 12 & 31  & 57 \\
PHP                            & 17 & 23  & 60 \\
Proj4                          & 7  & 43  & 50 \\
QuickJS                        & 29 & 28  & 43 \\
Solidity                       & 29 & 71  & 0 \\
SQL parser                     & 82 & 12  & 6 \\
Vorbis                         & 0  & 0   & 100 \\
\bottomrule
\end{tabular}
\end{table}

\paragraph{Answer to RQ2.}
Constraint-satisfaction failures are the largest aggregate outcome in both modes, but navigation also matters. Removing feedback reduces target-function reach from 75.2\% to 62.6\% and near-to-pass conversion from 48.4\% to 30.8\%, indicating that feedback supports both stages.

The trajectory audit supports this diagnosis: with feedback, successful runs use intermediate rewards to revise candidate-specific hypotheses, whereas open-loop runs can stop after an undetected near miss ($r{=}1.0$). Passing trajectories also sustain longer interaction, but tool volume alone does not explain success; these observations are descriptive because only a subset of jobs has complete readable trajectories.

\subsection{How Sensitive Are Results to Agent Configuration?}
\label{sec:ablation}

We isolate two evaluation choices, reasoning effort and agent scaffold, through targeted ablations. Each varies one factor while holding the remaining configuration fixed.

\subsubsection{Reasoning Effort}
\label{sec:ablation:effort}

We vary the reasoning effort of GLM-5.1 under Claude Code across low, medium, and high settings in both modes (Table~\ref{tab:ablation-effort}). In the Feedback-enabled mode, the pass rate rises monotonically, from 25.0\% at low to 28.3\% at medium and 38.3\% at high. In the Open-loop mode, high achieves 18.3\%, compared with 6.7\% at low and 3.3\% at medium. All effort settings complete 60 valid trials in each mode.

\begin{table}[H]
\centering
\caption{Reasoning-effort ablation; $n$ denotes valid trials.}
\label{tab:ablation-effort}
\preprinttableformat
\setlength{\tabcolsep}{8pt}
\begin{tabular}{@{}lrrrrr@{}}
\toprule
\textbf{Effort} & \textbf{$n$} & \textbf{Nav.} & \textbf{Constr.} & \textbf{Pass} & \textbf{Rate} \\
\midrule
\multicolumn{6}{@{}l}{\emph{Feedback-enabled}}\\
low     & 60 & 13 & 32 & 15 & 25.0\% \\
medium  & 60 & 10 & 33 & 17 & 28.3\% \\
high    & 60 & 14 & 23 & 23 & 38.3\% \\
\midrule
\multicolumn{6}{@{}l}{\emph{Open-loop}}\\
low     & 60 & 26 & 30 & 4  & 6.7\% \\
medium  & 60 & 27 & 31 & 2  & 3.3\% \\
high    & 60 & 23 & 26 & 11 & 18.3\% \\
\bottomrule
\end{tabular}
\end{table}

High effort changes the failure distribution differently across modes. With feedback, it produces 14 navigation failures, compared with 10 or 13 at the other settings, while increasing the number of passes to 23. Without feedback, it produces 11 passes, compared with four at low and two at medium. Reasoning effort is therefore not monotonically associated with performance across modes; its effect depends on the available feedback and should be interpreted cautiously.

\subsubsection{Scaffold Choice: Vendor vs.\ Neutral}
\label{sec:ablation:scaffold}

We compare two models under Terminus~2 and their vendor scaffolds in the Feedback-enabled mode (Table~\ref{tab:abl-scaffold}): Kimi Code for Kimi-K2.6 and Qwen Coder for Qwen3.7-Max. Kimi-K2.6 achieves 16.7\% under Kimi Code and 3.7\% under Terminus~2, while Qwen3.7-Max achieves 13.3\% under Qwen Coder and 0.0\% under Terminus~2. The consistent direction shows that scaffold choice affects measured performance for these models, but two comparisons are insufficient for broader generalization.
\begin{table}[H]
\centering
\caption{Pass rates by scaffold.}
\label{tab:abl-scaffold}
\preprinttableformat
\begin{tabular}{@{}lrr@{}}
\toprule
\textbf{Model} & \textbf{Terminus~2} & \textbf{Vendor scaffold} \\
\midrule
Kimi-K2.6        & 3.7\% & 16.7\% \\
Qwen3.7-Max      & 0.0\% & 13.3\% \\
\bottomrule
\end{tabular}
\end{table}
We further audit the recorded trajectories (Table~\ref{tab:abl-scaffold-traj}), reporting median tool calls and verifier queries per available trajectory; this analysis is descriptive because some jobs lack complete trajectories.
\begin{table}[H]
\centering
\caption{Trajectory activity by scaffold.}
\label{tab:abl-scaffold-traj}
\preprinttableformat
\begin{tabular}{@{}llrrr@{}}
\toprule
\textbf{Model} & \textbf{Scaffold} & \textbf{Traj.} & \textbf{Tool calls} & \textbf{Verifier queries} \\
\midrule
Kimi-K2.6 & Kimi Code  & 60 & 68.0  & 4.0  \\
          & Terminus~2 & 54 & 242.5 & 25.5 \\
Qwen3.7-Max & Qwen Coder & 57 & 76.0 & 9.0 \\
            & Terminus~2  & 60 & 54.5 & 4.0 \\
\bottomrule
\end{tabular}
\end{table}
Tool volume alone does not explain the scaffold gap: for Kimi-K2.6, Terminus~2 makes more tool calls and verifier queries than Kimi Code, yet achieves a lower pass rate. The vendor scaffolds yield 10 and 8 branch-satisfying inputs for Kimi-K2.6 and Qwen3.7-Max, compared with 1 and 0 under Terminus~2. Thus, scaffold design affects how interactions are converted into target-branch satisfaction, not merely how many actions an agent performs. Measured performance is sensitive to both reasoning effort and scaffold choice: the effort ranking changes with feedback availability, while vendor scaffolds outperform Terminus~2 in both evaluated comparisons. Model comparisons should therefore control and report the task mode, reasoning effort, and scaffold; the scaffold result remains limited to two models.

\section{Discussion and Threats to Validity}
\label{sec:discussion}
\label{sec:threats}
 
\subsection{Discussion}
\label{sec:discussion:impl}
 
The results identify two distinct bottlenecks: reaching the target function and satisfying its branch constraint. Feedback improves both target-function reach (62.6\% to 75.2\%) and near-to-pass conversion (30.8\% to 48.4\%), while model and scaffold choices change the magnitude of the pass-rate difference. Evaluation should therefore report both stages rather than attributing all failures to either navigation or refinement alone.
 
Constraint-satisfaction failures remain the largest aggregate outcome, motivating feedback that identifies which sub-condition of a guard remains unsatisfied. At the same time, the navigation gap shows that feedback must also help agents reach the relevant code. Our scaffold and reasoning-effort findings (Section~\ref{sec:ablation}) further show that leaderboards should disclose the harness, mode, and effort setting before ranking models.
 
\subsection{Threats to Validity}
\label{sec:threats:ttv}
 
SWE-Test measures vulnerability discovery through single branch inversion rather than full exploit construction. We chose this target because it is deterministically checkable and can be pinpointed to one of the three abilities, and the four-level reward separates reaching the target function from satisfying the target branch constraint rather than collapsing both into pass/fail; still, it is a proxy, so results describe the comprehension, correction, and exploration stack rather than end-to-end exploit construction. In its fixed-target modes, the benchmark covers 60 tasks over 16 C/C++ OSS-Fuzz~\cite{serebryany2017ossfuzz} targets and 13 domains, concentrated in a few high-task-count domains such as PHP, SQL, XML, and fonts, so findings may not transfer to other languages or input formats; Online Arena adds 6 further programs and 2 further domains not in this fixed-target pool, for 22 distinct programs and 15 domains benchmark-wide (Section~\ref{sec:bench:overview}). Every task is built from already-known, fuzzer-discovered branches in public projects, so the released artifacts contain no novel exploit or undisclosed vulnerability.
 
Two factors limit how directly we can compare models. Scaffold choice can materially change pass rate, so rankings must hold the scaffold fixed. Models also use default reasoning effort rather than a matched level, and Section~\ref{sec:ablation:effort} shows that the effort settings shift GLM-5.1's feedback-enabled pass rate from 25.0\% to 38.3\%. Finally, some configurations have fewer than 60 valid trials; all reported pass rates exclude unscored trials, so comparisons with different denominators should be interpreted cautiously.

\section{Conclusion}
\label{sec:conclusion}

We introduce SWE-Test, a benchmark that evaluates vulnerability-discovery agents through input prediction on real C/C++ programs. Its Open-loop, Feedback-enabled, and Online Arena modes separate code comprehension, feedback-driven correction, and path exploration. Across 15 default-effort model-scaffold configurations, the best feedback-enabled pass rate is 55.0\%. Among seven paired Claude Code configurations, feedback raises the mean pass rate from 19.3\% to 36.4\% and improves both target-function reach and target-constraint satisfaction on average. Scaffold choice and reasoning effort also materially affect measured performance, making uncontrolled model rankings difficult to interpret. We release 60 Open-loop and Feedback-enabled tasks across 16 programs, the Online Arena, and the evaluation harness to support reproducible analysis of where vulnerability-discovery agents fail.

\bibliographystyle{colm2024_conference}
\bibliography{references}

\end{document}

%% file: author_info.tex
\newcommand{\paperauthors}{%
\small
\href{https://openreview.net/profile?id=~Shi_Yuanxiang1}{Yuanxiang Shi}\textsuperscript{*},
\href{https://openreview.net/profile?id=~Jiayi_Lin15}{Jiayi Lin}\textsuperscript{*},
\href{https://openreview.net/profile?id=~Xuanyong_Lin1}{Xuanyong Lin}\textsuperscript{*},
\href{https://openreview.net/profile?id=~Liangcai_Su1}{Liangcai Su},
\href{https://openreview.net/profile?id=~Duan_Yeheng1}{Yeheng Duan},
\href{https://openreview.net/profile?id=~Wei_Wang2}{Wei Wang},
\href{https://openreview.net/profile?id=~Qi_Han11}{Qi Han},
\href{https://openreview.net/profile?id=~Bing_Zhao7}{Bing Zhao},
\href{https://openreview.net/profile?id=~HU_WEI4}{Wei Hu},
\href{https://openreview.net/profile?id=~Xander_Xu1}{Xander Xu}\textsuperscript{\textdagger},
\href{https://openreview.net/profile?id=~Chenxiong_Qian1}{Chenxiong Qian}\textsuperscript{\textdagger}\\[0.35em]
{\fontsize{8.5pt}{9.5pt}\selectfont
\textsuperscript{*}\,Equal contribution.\quad
\textsuperscript{\textdagger}\,Corresponding authors.}\\[0.15em]
}